\def\simlt{\lower.5ex\hbox{$\; \buildrel < \over \sim \;$}}
\def\simgt{\lower.5ex\hbox{$\; \buildrel > \over \sim \;$}}
\RequirePackage{lineno} 
 \documentclass[preprint2]{emulateapj}

\newcommand{\myemail}{mrl@gps.caltech.edu}
\slugcomment{Submitted to the Astrophysical Journal }
\shorttitle{Line et al.}
\shortauthors{Line et al.}
\begin{document}
\title{A Near-Infrared Transmission Spectrum for the Warm Saturn HAT-P-12b}
\author{Michael R. Line}
\affil{Division of Geological and Planetary Sciences, California Institute of Technology, Pasadena, CA 91125}
\author{Heather Knutson}
\affil{Division of Geological and Planetary Sciences, California Institute of Technology, Pasadena, CA 91125}
\author{Drake Deming}
\affil{Department of Astronomy, University of Maryland, College Park, MD 20742}
\author{Ashlee Wilkins}
\affil{Department of Astronomy, University of Maryland, College Park, MD 20742}
\author{Jean-Michel Desert}
\affil{Division of Geological and Planetary Sciences, California Institute of Technology, Pasadena, CA 91125}

\email{mrl@gps.caltech.edu}
\altaffiltext{1}{Correspondence to be directed to \myemail}
\begin{abstract}
 We present a HST WFC3 transmission spectrum for the transiting exoplanet HAT-P-12b.  This warm (1000 K) sub-Saturn-mass planet has a smaller mass and a lower temperature than the hot-Jupiters that have been studied so far.  We find that the planet's measured transmission spectrum lacks the expected water absorption feature for a hydrogen-dominated atmosphere, and is instead best-described by a model with high-altitude clouds.  Using a frequentist hypothesis testing procedure, we can rule out a hydrogen-dominated cloud free atmosphere to 4.9$\sigma$.  When combined with other recent WFC3 studies, our observations suggest that clouds may be common in exo-planetary atmospheres. 
 

\end{abstract}


\section{Introduction}
Observations of transiting planets provide an invaluable window into the nature of exoplanet atmospheres.   Specifically, measuring the wavelength dependent transit depth allows us to determine the presence of absorbing gases (Charbonneau et al. 2002), atmospheric scale height (Miller-Ricci, Seager \& Sasselov 2009), and the presence of high-altitude hazes (Pont et al. 2008).  Recently, the HST Wide Field Camera-3 (WFC3) has been used for both emission and transmission spectroscopy between 1.1 and 1.8 microns (Berta et al. 2012; Swain et al. 2013; Deming et al. 2013; Wilkins et al. 2013; Huitson et al. 2013; Stevenson et al. 2013).  This spectroscopic region contains strong absorption features due to water, and weaker features due to methane, carbon-monoxide, and carbon dioxide.  Furthermore, Swain et al. (2013) suggest the possibility of absorption due to metal oxides at shorter wavelengths.  Determining the relative amplitudes of absorption features in this window can constrain the atmospheric mean molecular weight, allowing us to infer the atmospheric metallicity and the dominant atmospheric constituents.  Additionally, observations indicating a lack of absorption features over this spectral region may suggest the presence of high altitude clouds or hazes.  Hazes and clouds can be due to either equilibrium condensates or photochemically produced Titan-like hazes.

To date, WFC3 transmission observations have been reported for several planets including WASP 12b, GJ 1214b, HD 209458b,WASP-19 and XO-1b.   In this investigation we examine the WFC3 transmission spectrum of the warm Saturn HAT-12b.  HAT-P-12b was discovered with the HATNet (Bakos et al. 2006) survey and found to be a low density (0.295 g cm$^{-3}$) sub-Saturn mass planet orbiting a metal poor, 4650 K, 0.701$R_{\odot}$ star (Hartman et al. 2009).   The planet is in a 3.2 day (0.084 AU) orbit and has a radius of 0.96$R_{J}$ and mass of 0.21$M_{J}$.  The equilibrium temperature is 965 K assuming full redistribution and zero albedo. Miller \& Fortney et al. (2011) demonstrated that the mass and radius of this planet are consistent with an H/He (76\% by mass) planet with a core mass of 17 M$_{\oplus}$.  HAT-P-12b's low density and relatively bright primary make it a favorable target for transmission spectroscopy, allowing us to explore the properties of exoplanetary atmospheres in this low-temperature regime. In this paper we present the first measurements of this planet's transmission spectrum.  We first describe the observations followed by a discussion of the data reduction approach and a simple modeling analysis of plausible scenarios for the planet's atmospheric composition.

\section{Observations} 
HAT-P-12b was observed on May 29th 2011 4:08:48 - 9:42:56 UT using the G141 grism of WFC3 in one visit as part as HST program 12181 (PI D. Deming).  We obtained 111 images over the course of 4 orbits.   Each exposure spans 12.79 seconds.  These observations were made before the implementation of the spatial scan mode (Deming et al. 2013), and therefore use a fixed pointing on the array.

\section{Data Reduction}
\subsection{Extracting the White Light Curve}
Our data reduction approach is similar to that of Berta et al. 2012, Deming et al. 2013, and Wilkins et al. 2013.   We first use the direct image of the star to set the reference point on the image by fitting the point spread function with a 2-D Gaussian.  The x-position is needed for the wavelength calibration and the y-position sets the center of the point spread function (PSF) for the extraction box.  The extraction box is fixed in the y-direction for the remaining images.  For each image we use an extraction box size of 25 pixels (roughly 7 spatial full-widths at half max of the PSF) in the spatial direction and 150 pixels in the wavelength direction centered about the 1st order spectrum.  The images are subject to standard processing techniques.  We use a 5$\sigma$ median filter in time to remove cosmic rays in each pixel (0.007$\%$ of all pixels).  The pixels affected by cosmic rays are replaced with the median value from all other images in the time series.  We also flat field and subtract the wavelength dependent sky background.  The sky background is estimated using a box that is the same size as the first order spectral extraction box but offset by 100 pixels in the y (cross-disperse) direction.  The sky image is also flat fielded before extracting the sky background.  

\begin{figure}
  \centering
    \includegraphics[width=0.45\textwidth]{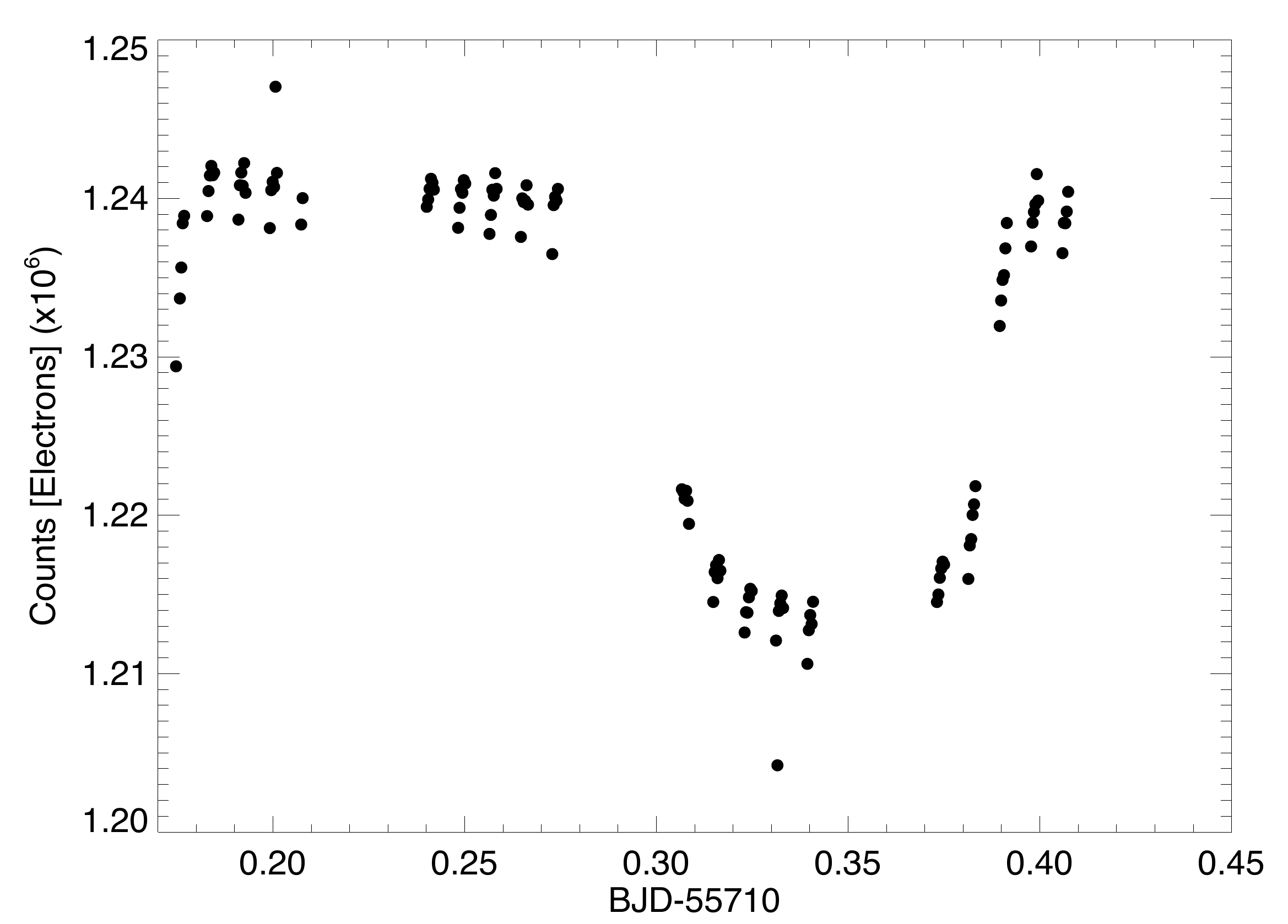}
     \caption{\label{fig:figure1} Raw light curve.  The total flux for each image is given in electrons.  There are 4 orbits each of which contain 5 buffer dumps. After each buffer dump there is a residual charge build up that produces hook-like features.  These features are reset after each buffer dump}
\end{figure}

In order to construct the white light transit curve we sum all the pixel values in electrons within each of the extracted images.    Upon computing the number of counts for each image we can construct the raw light curve (Figure \ref{fig:figure1}).   The Julian Date (JD) given in the header is converted to Barycentric Julian Date (BJD) using the IDL routines of  Eastman, Siverd \& Gaudi (2010).  We do not include the first orbit in our analysis because of unrepeatable systematics due to the thermal variations in the telescope that occur after a new target acquisition.  Some basic features to notice in Figure \ref{fig:figure1} are, first, the 4 separate orbits and second, within each orbit there are 5 batches due to the buffer dump, each of which contain either 5 or 6 individual exposures.  In each batch there is a hook like feature that is due to a build up of charge on the detector (Wilkins et al. 2013).  After each buffer dump the residual charge is automatically reset.  

We also find an outlier in our white-light curve near the center of the transit.  This outlier also has a consistently low value in the light curves for our individual bands.  We exclude this point from our subsequent analysis.  Neglecting this outlier does not affect the shape of the transmission spectrum.  We considered whether errors in the flat field or sky background or missed cosmic rays  could explain the outlier, but cannot identify any clear explanation.

\subsection{Extracting the 1st Order Spectrum}
We estimate the wavelength-dependent transit depth by subdividing the first order spectrum image into wavelength bins.   The pixel values within each bin are summed to obtain the total number of counts within that bin.   We choose a bin size of 5 pixels corresponding to a spectral resolution of 23 nm (Berta et al. 2012).  This produces 30 spectral channels across the first order spectrum.   The slope of the first order spectrum on the CCD does not change by more than one pixel across all 150 pixels, hence we can sum the pixels column-by-column.  From tracking stellar features from image to image we find that horizontal jitter in the first order spectrum is virtually non existent.   A wavelength calibration, or a mapping of the horizontal pixel number onto wavelength is preformed using the wavelength calibration function from the STScI pipeline (Kuntschner et al. 2009) with updated calibration coefficients from Wilkins et al. 2013.  The resulting spectrum spans from 1.037 $\mu$m to 1.721 $\mu$m.

\section{Parameter Estimation}
We model the eclipse with the Mandel \& Agol (2002) IDL routines.  This parameterization is governed by four free parameters:  the ratio of the semi-major axis to the stellar radius ($a/R_{*}$), the inclination ($i$), the center-of-transit time ($t_0$), and the planet radius to stellar radius ($R_{p}/R_{*}$). We also include non-linear limb darkening parametrized with four coefficients.  The limb darkening coefficients are determined by fitting a non-linear limb darkening parameterization (equation 6 in Claret 2000) to an intensity weighted ATLAS stellar spectrum (T$_{*}$=4650 K, log$g$=4.6, [M/H]=-0.3) over the appropriate wavelength range.  We derive both the white-light limb darkening coefficients and the coefficients for each separate wavelength bin.  

In addition to modeling the eclipse depth, we also model the detector systematics.  We use the ``model-ramp" parameterization described in Berta et al. 2012 given by
\begin{equation}\label{eq:equation1}
\frac{F_{obs}}{F_{cor}}=(C+Vt_{vis}+Bt_{orb})(1-Re^{t_{batch}/\tau})
\end{equation}

This model adds an additional 5 free parameters ($C, V, B, R, \tau$) to our total set of parameters.  This ``model-ramp" parameterization accounts for a visit long slope ($V$), slope with-in each orbit ($B$), vertical offset ($C$), and an exponential model for hook ($R, \tau$).  The resulting array from this equation is multiplied by the model light curve  from the Mandel \& Agol 2002 routine.  We have chosen the ``model-ramp" approach rather than the divide-out-of-transit approach (Berta et al. 2012) to account for systematics that are not consistent from orbit to orbit (Wilkins et al. 2013).  Additionally, the ``model-ramp" procedure allows us to include one orbit of out-of-transit data (the second orbit in figure \ref{fig:figure1}), whereas the divide-out-of-transit approach utilizes this orbit to correct the in-transit orbits.  Including this out-of-transit orbit provides a better constrained base-line for the light curve model.   

\begin{figure}
  \centering
    \includegraphics[width=0.45\textwidth]{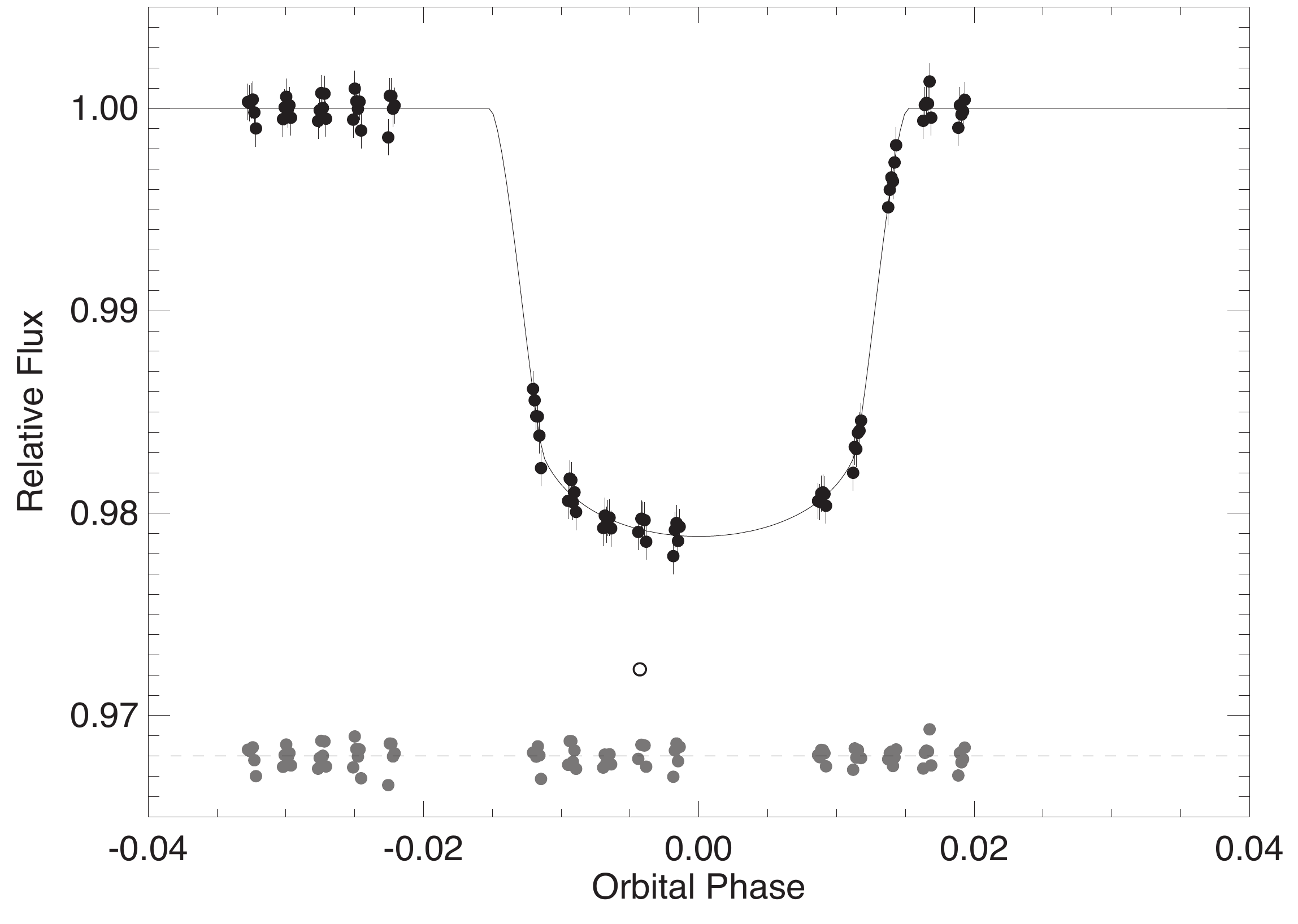}
     \caption{\label{fig:figure2}Transit model fit to the white light curve.  The systematics (see equation 1) are removed from the data (round points with error bars) in this plot.  The solid curve is the best fit light curve model.  The gray circles are the residuals.  The hollow circle is an outlier point that we exclude from the fitting. }
\end{figure}

We use the IDL MPFIT (Markwardt 2009) Levenbergh-Markquardt curve fitting routine to find the optimal set of parameter values for the white light curve and each wavelength bin.  The wavelength independent parameters ($a/R_{*}$, $i$, and  $t_0$) are determined from the best-fit white-light curve (Table \ref{tab:table1}).  They are fixed at the best-fit values when fitting the light curves for each wavelength bin.  We first fit the light curves without error bars.  We then compute the root-mean-square (RMS) of the residuals between the data and the best fit determined by MPFIT.  This RMS value is then used as the actual uncertainty on each point.  This is done for both the white-light eclipse and the eclipses in each spectral channel.  Upon deriving these RMS error bars, the fit is performed again to obtain the nominal set of model parameters and uncertainties.  The final uncertainties on all parameters are the Gaussian uncertainties derived from the MPFIT covariance matrix.  It has been shown in Berta et al. (2012) that the MPFIT covariance derived parameter uncertainties are in good agreement with those derived from Markov-chain Monte Carlo approaches for this type of data.   We also performed a prayer-bead (Gillon 2009; Carter et al. 2009) analysis to explore the effects of time-correlated noise on the parameter uncertainties and generally find that they are also in good agreement with the MPFIT uncertainties.  We do not use our prayer-bead analysis for the final errors because of the sparsity in the number of data points sampling the light curve which leads to large uncertainties in the estimated errors.   Figure \ref{fig:figure2} shows the resultant fit to the white light transit curve with the systematics (equation \ref{eq:equation1}) divided out.

\begin{figure}
  \centering
    \includegraphics[width=0.45\textwidth, angle=0]{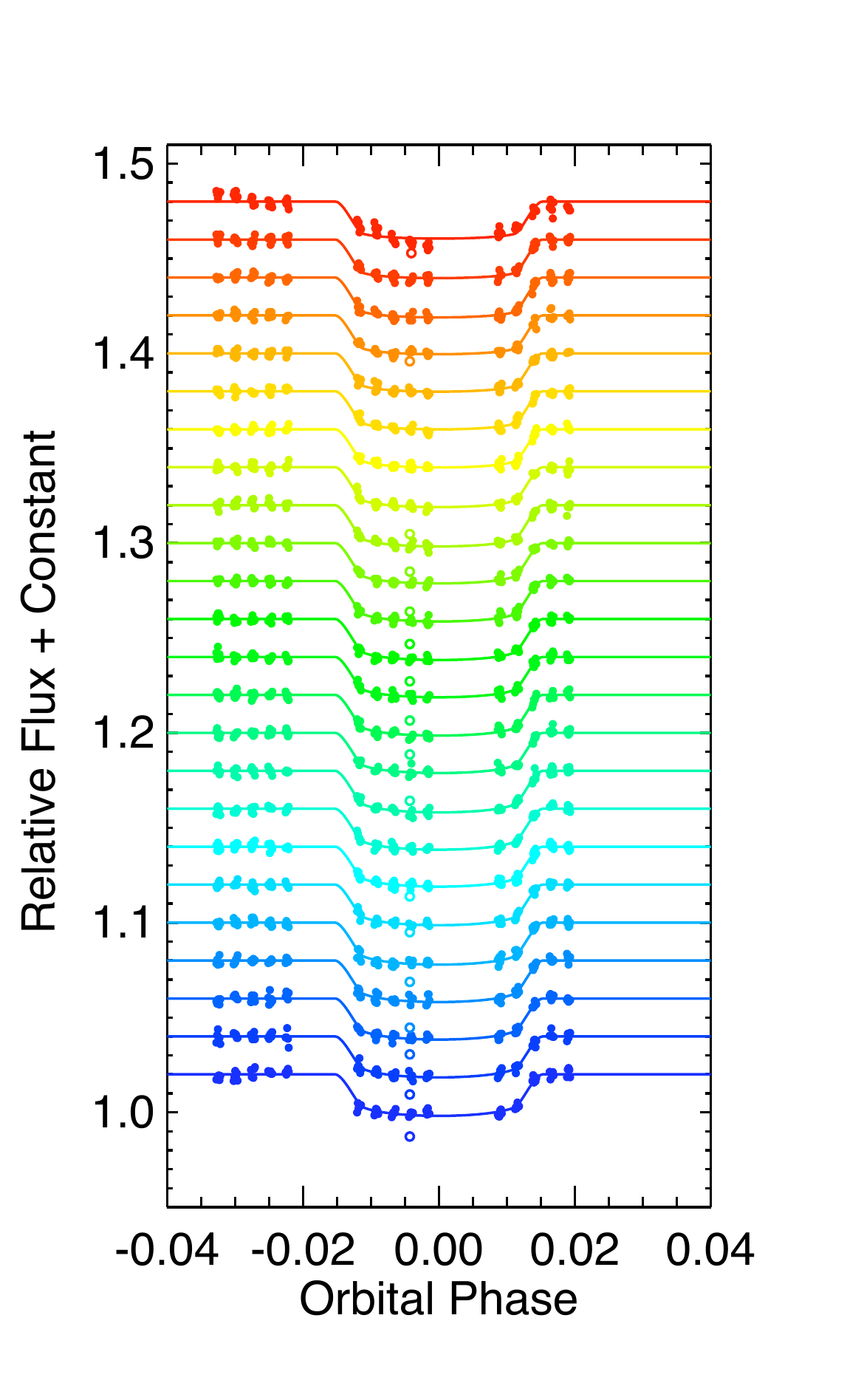}
     \caption{\label{fig:figure3}Transit model fit to each spectral bin.  The systematics are removed from the data (round points with error bars).  The solid curves are the best fit light curve models for each bin.  Transit eclipse depths for the shorter wavelengths are denoted by blue near the bottom and the longer wavelengths are shown in red near the top.  The hollow circles are the outlier point that we exclude from the fits.  }
\end{figure}

\begin{figure}
  \centering
    \includegraphics[width=0.45\textwidth, angle=0]{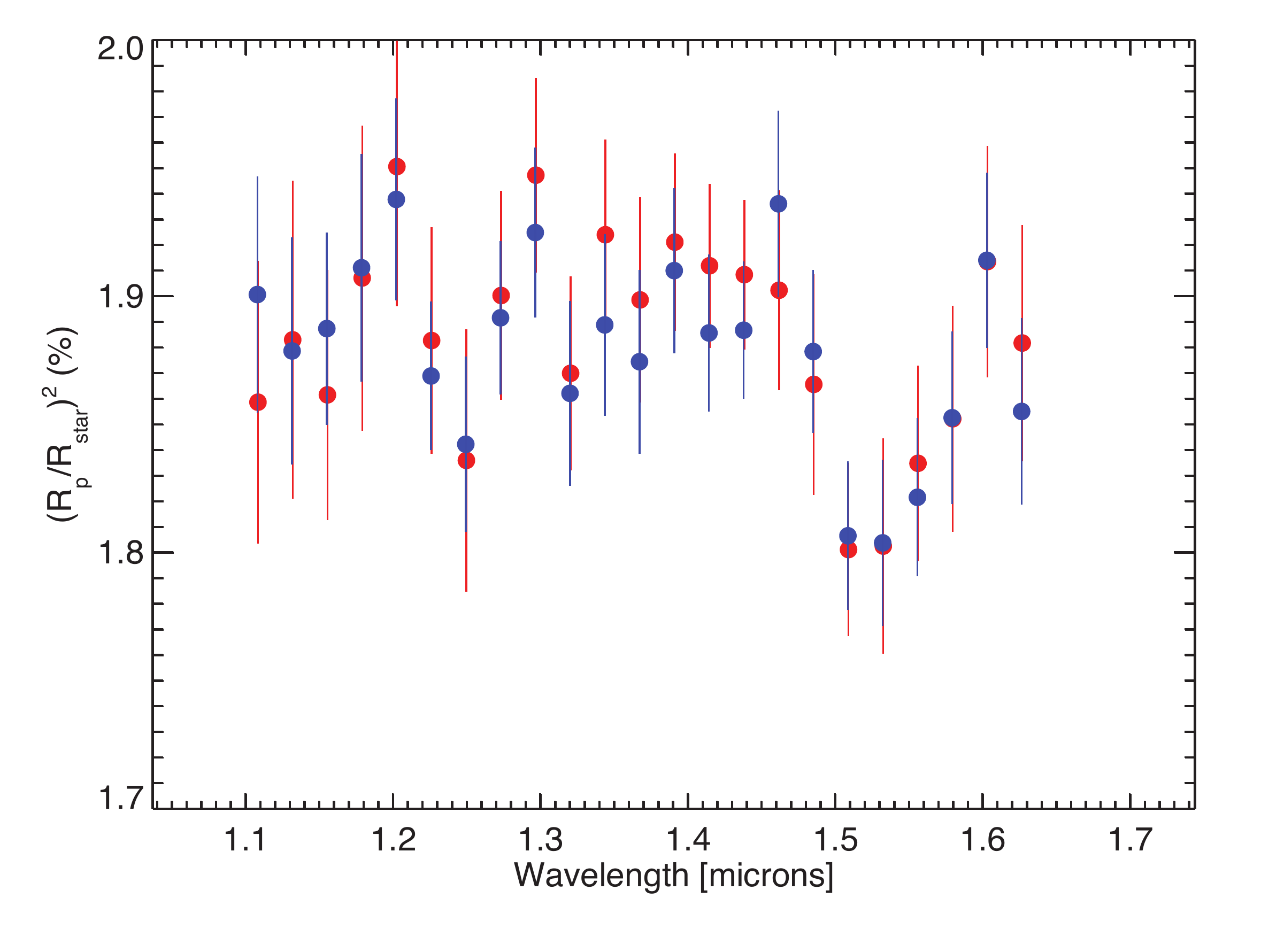}
     \caption{\label{fig:figure4} Transmission spectrum derived under different different detecter systematic model assumptions.  The red spectrum results from fitting for the detector systematics within each wavelength bin.  The blue spectrum is obtained by using the white-light curve derived residuals as the systematic structure for each wavelength bin.  The results are consistent.    }
\end{figure}

In addition to fitting for the systematics within each wavelength bin we also fit each wavelength bin using the ``divide-white" approach which uses fixed detector systematics derived from the white-light transit (Stevenson et al. 2012; Sing et al. 2013).   We first fit the white light curve as above with the systematics included.  We then subtract the best fit transit light curve from the data leaving a residual vector consisting only of the white-light instrument effects.  When fitting the light curves in each wavelength channel we then take this residual systematics vector, multiply it by a scale factor and add an offset.  This new vector is then added to the transit light curve model for that bandpass.  The free parameters when fitting each spectral bin are the eclipse depth, the residual systematics scale factor, and a constant offset for the systematics vector.  This method offers an advantage over the parameterized approach by reducing the number of free parameters required to fit the individual bandpasses.  It also avoids the need to assume a functional form for the systmatic noise, and is therefore more general than the previous approach.  Its primary limitation is the assumption that the visit-long linear trend, the ramp timescale, and the orbit-long linear trends are all independent of wavelength.      Figure \ref{fig:figure3} shows the eclipses in each wavelength channel with the systmatics removed.   Figure \ref{fig:figure4} compares the resultant transmission spectrum from each approach.  The two approaches produce consistent results, but the ``divide-white" fit has modestly reduce the uncertainties on the wavelength-dependent transit depth.  For the remainder of the analysis we focus on the transmission spectrum  (Table \ref{tab:table2}) derived from the ``divide-white" approach.   The noise per channel is on average, 1.3 times the photon noise.   The effective resolving power of the spectrum is 60 at 1.4 $\mu$m and the effective signal-to-noise per wavelength channel is approximately 50.  

We also explored the effects that the uncertainties in $a/R_{*}$, $i$, and  $t_0$ derived from the white light curve have on the spectrum.  We find that they simply result in a wavelength independent vertical shift, similar to those results found by Berta et al. 2012.   

\begin{center}
\begin{deluxetable*}{ccccc}
\tablecolumns{6} 
\tablewidth{0pt}  
\tablecaption{\label{tab:table1}White light derived parameters.  The center of transit times, $t_0$, for Hartman et al. (2009) and Sada et al. (2012) are adjust to our epoch (402 orbits later).
      }
\tablehead{\colhead{Parameter }			&
		\colhead {This Work }                                    &
		\colhead {Hartman et al. 2009 }                                    &
				\colhead {Sada et al. 2012 }                                    &
	                           }
\startdata
$a/R_{*}$   &   11.6$^{+0.39}_{-0.39}$	& 11.8$^{+0.15}_{-0.21}$  & 11.2$^{+0.45}_{-0.69}$ 	\\
$i (^{\circ})$	&	88.7$^{+0.62}_{-0.62}$ 	& 89.0$^{+0.4}_{-0.4}$  &	88.5$^{+0.99}_{-0.93}$ \\
$t_0 (BJD)$    &     2455710.8453$^{+0.00022}_{-0.00022}$ 	& 2455710.9001 $^{+0.00020}_{-0.00020}$  &	2455710.89826 $^{+0.00020}_{-0.00020}$\\
$R_{p}/R_{*}$ & 0.137 $^{+0.0011}_{-0.0011}$ 	& 0.141$^{+0.0013}_{-0.0013}$  & 	0.140$^{+0.0026}_{-0.0026}$\\
\enddata
\end{deluxetable*}
\end{center}	

\begin{center}
\begin{deluxetable}{cccc}
\tablecolumns{6} 
\tablewidth{0pt}  
\tablecaption{\label{tab:table2}Derived transmission spectrum.
      }
\tablehead{\colhead{Wavelength [$\mu$m] }			&
		\colhead {$(R_{p}/R_{*})^2$ }                                    &
				\colhead {Uncertainty}                                    &
	                           }
\startdata
1.108  &   0.01901  &   0.00046  \\
1.132  &   0.01879  &   0.00044  \\
1.155  &   0.01887  &   0.00038  \\
1.179  &   0.01911  &   0.00044  \\
1.202  &   0.01938  &   0.00039  \\
1.226  &   0.01869  &   0.00029  \\
1.250  &   0.01842  &   0.00034  \\
1.273  &   0.01892  &   0.00030  \\
1.297  &   0.01925  &   0.00033  \\
1.320  &   0.01862  &   0.00036  \\
1.344  &   0.01889  &   0.00036  \\
1.367  &   0.01874  &   0.00036  \\
1.391  &   0.01910  &   0.00032  \\
1.414  &   0.01886  &   0.00031  \\
1.438  &   0.01887  &   0.00027  \\
1.462  &   0.01936  &   0.00036  \\
1.485  &   0.01878  &   0.00032  \\
1.509  &   0.01806  &   0.00029  \\
1.532  &   0.01804  &   0.00032  \\
1.556  &   0.01821  &   0.00031  \\
1.579  &   0.01853  &   0.00034  \\
1.603  &   0.01914  &   0.00034  \\
1.627  &   0.01855  &   0.00036  \\
\enddata
\end{deluxetable}
\end{center}


\section{Interpretation}
Transmission spectra are useful in determining molecular abundances, atmospheric mean molecular weight, and the presence of high-altitude clouds.  Due to the limited wavelength coverage and SNR of our data, we do not attempt to perform a rigorous atmospheric retrieval (Madhusudhan \& Seager 2009; Madhusudhan et al. 2011; Lee et al. 2012; Line et al. 2012;  2013;  Benneke \& Seager 2012; 2013).   However, this spectral region should show strong H$_2$O absorption if the elemental abundances are near solar values.  If we do not see the water vapor absorption feature as has been detected in a variety of other planets observed with WFC3, then we might infer the presence of a high-altitude cloud that effectively damps the amplitude of the absorption features.  Another possibility to explain the lack of water absorption is the lack of any molecular absorbers or a high carbon-to-oxygen ratio.  In this analysis we consider three scenarios: a solar composition clear atmosphere,  a solar composition atmosphere with an opaque gray cloud at 1 mbar, and an atmosphere devoid molecular absorption other than continuum.    It is not unreasonable to assume the presence of clouds given the likely possibility of several equilibrium condensates (e.g., Morley et al. 2013) and possible photochemical aerosols in the pressure-temperature regime of the upper atmosphere.  We need not concern ourselves with the notion of a high mean molecular weight atmosphere due to the planets extraordinarily low density which requires a thick H/He atmosphere (Hartman et al. 2009; Miller \& Fortney 2011).    

\begin{figure}
  \centering
    \includegraphics[width=0.45\textwidth]{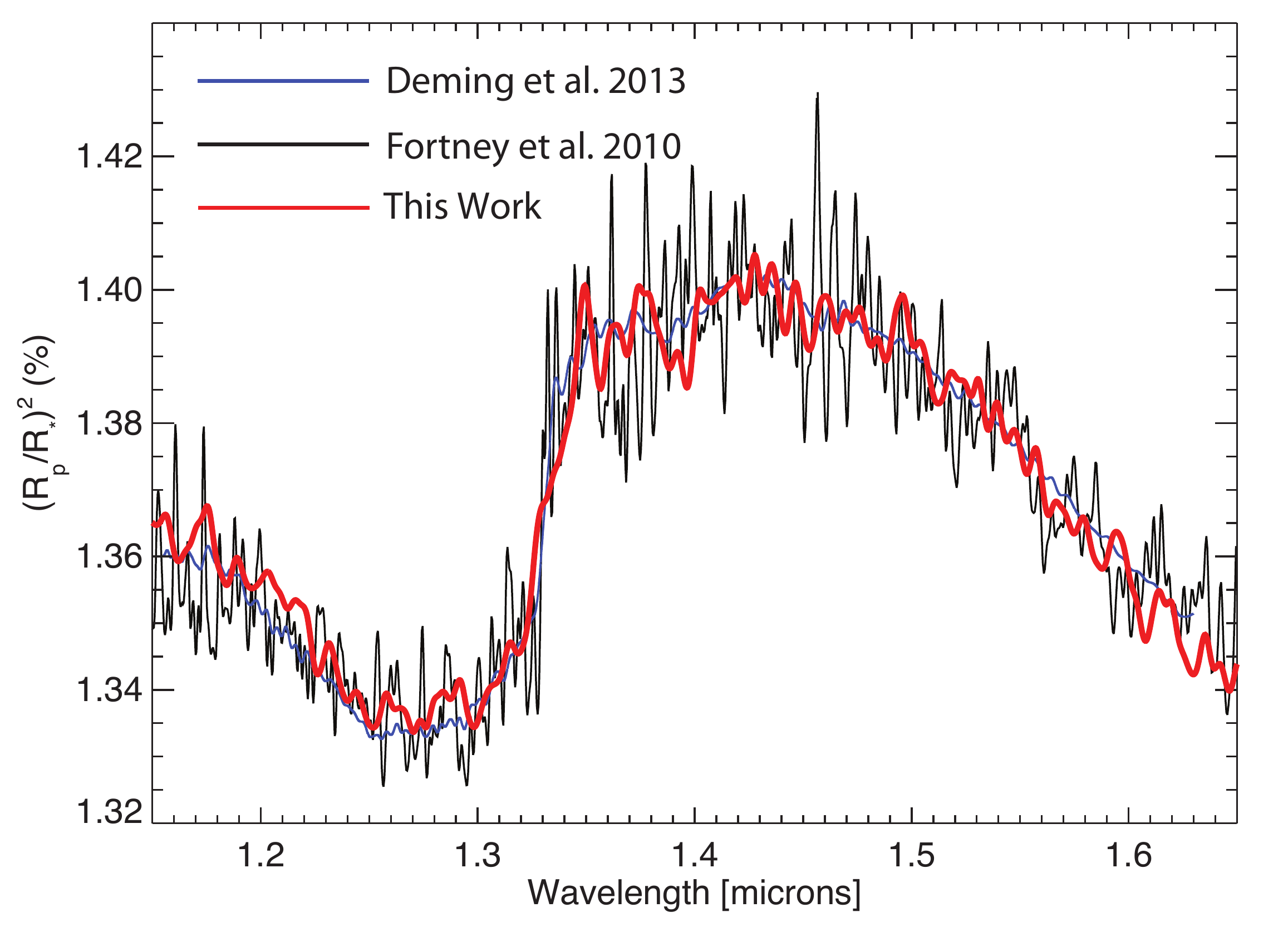}
     \caption{\label{fig:figure5}Transmission code validation.  We compare results of our model with those of Deming et al. 2013 and Fortney et al. 2010 based on HD209458b planetary parameters.  In this comparison the planetary radius is defined to be 1.25$R_{J}$ at 10 bars, the stellar radius is 1.148$R_{\odot}$, and a planet gravity of 10 ms$^{-2}$.  We assume a 90 layer atmosphere starting at 10 bars extending to 10$^{-10}$ bars evenly spaced in log(pressure).  The atmosphere is assumed to be isothermal at 1500 K with mole fractions of 0.85, 0.15, and $4.5\times10^{-4}$ for H$_2$, He, and H$_2$O, respectively.  We assume no other absorbing gases.      }
\end{figure}

We have constructed a radiative transfer model that computes a transmission spectrum given the molecular abundances, temperature structure, cloud levels etc.  The model divides the planet up into annuli and computes the integrated slant optical depth and transmittance along each tangent height. The effective eclipse depth of the planet at each wavelength is then computed by integrating the slant transmittance profile using equation 11 in Brown (2001).  The molecular cross sections we use here are described in Line et al. (2013).  We have validated our model against those presented in Figure 12 of Deming et al. (2013), reproduced here in Figure \ref{fig:figure5}.  

\begin{figure}
  \centering
    \includegraphics[width=0.45\textwidth]{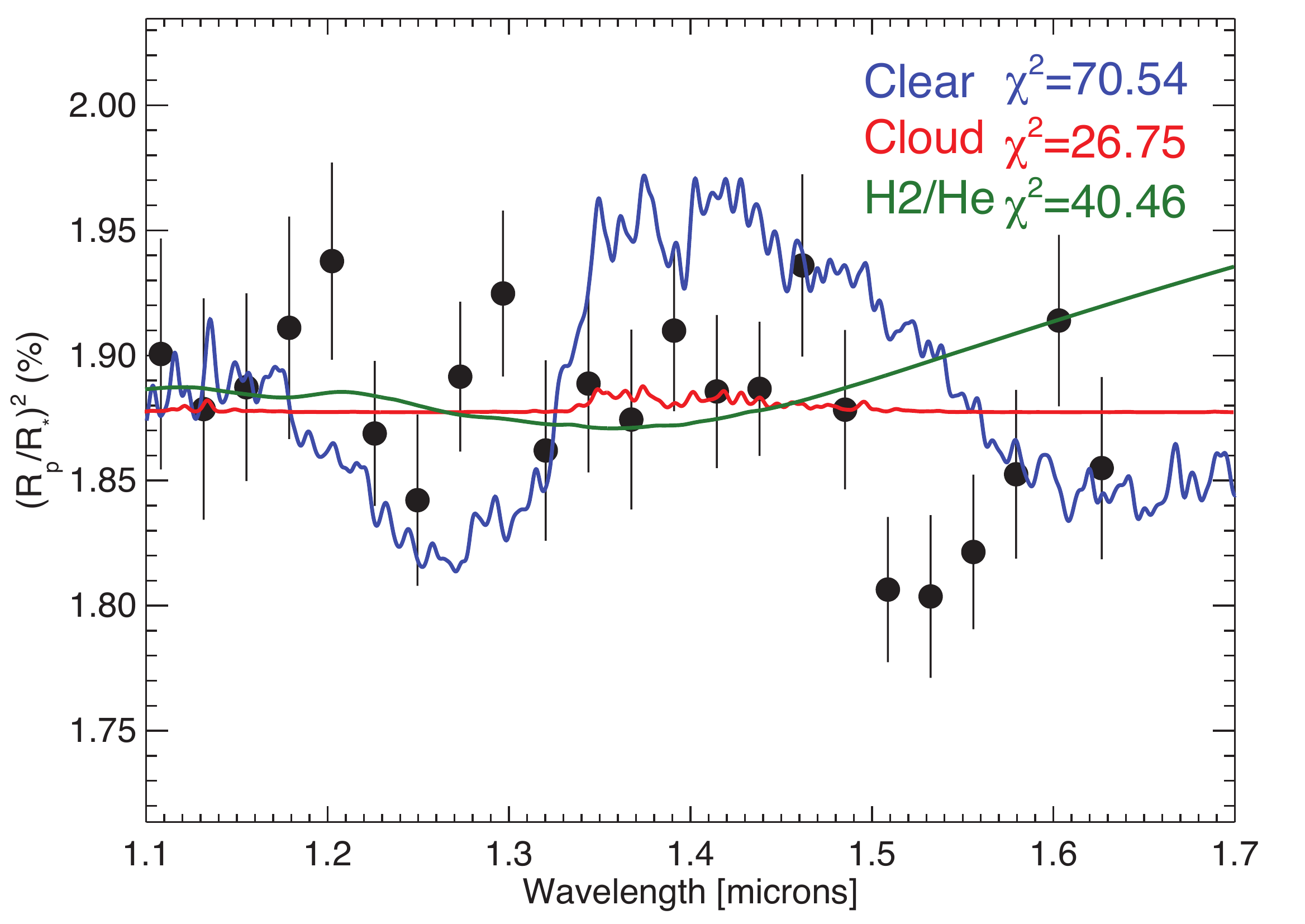}
     \caption{\label{fig:figure6}Transmission spectrum models compared with the data.  There are 3 models shown here.  The blue is a solar composition atmosphere, red is solar composition with a 1 mbar opaque cloud, and green is an atmosphere devoid of molecular absorbers other than continuum.  Upon using a ``frequentist" hypothesis testing procedure we can rule out the solar composition and molecule free atmospheres, but cannot rule out a cloudy atmosphere (see text).    }
\end{figure}

We generate solar composition thermochemical equilibrium mixing ratio profiles using the NASA Chemical Equilibrium with Applications Model (Gordon \& Mcbride 1996).  We assume a generic irradiated gas giant temperature profile using an analytic parameterization (Guillot 2010; Heng et al. 2012; Robinson \& Catling 2012).  The sensitivity of the transmission spectrum to the detailed structure in the temperature profile is minimal.  In order to correctly match the model spectra to the data, we shift the model spectrum vertically such that its average $(R_{p}/R_{*})^2$ is equal to that of the data.  This is equivalent to adjusting the pressure level at which the planetary radius is defined.    We integrate the high resolution model spectrum over each wavelength channel to the data points when undergoing the model comparison. The clear model is shown as the blue spectrum in Figure \ref{fig:figure6} and the cloudy spectrum is shown in red.  

We undergo a ``frequentist" hypothesis testing procedure (Gregory 2005 pp. 163-166)  to determine if we can rule out either of these three scenarios.  We treat each scenario independently as a null hypothesis.  If we can rule out the null hypothesis, then that suggests some other explanation must be needed.   For each of the three scenarios we first compute $\chi^2$.   We then compute the p-value, or the value that describes the probability of drawing a $\chi^2$ value larger than the given value for a repeated set of measurements given the same model.  This p-value can then be converted into a confidence interval in terms of how well we think we can rule out a given model.  From the two $\chi^2$ values in Figure \ref{fig:figure4} we can rule out a clear atmosphere to 4.9$\sigma$, a cloudy atmosphere to only 1.1$\sigma$, and a water free atmosphere to 3$\sigma$.  These results suggest that a cloudy atmosphere is the most physically plausible scenario.

\section{Discussion \& Conclusions}
The HST WFC3 is a powerful tool for studying the atmospheres of extrasolar planets.  We have performed an analysis on the low density cool exoplanet atmosphere of HAT-P-12b.  We found using a hypothesis testing procedure that a solar composition, clear atmosphere and a water free atmosphere are inconsistent with the data whereas a cloudy scenario is in good agreement.    It is physically plausible that clouds can exist at high altitudes in exo-planetary atmospheres.  According to Morley et al. (2013) there are three possible equilibrium condensates, Na$_2$S, ZnS, and KCl, in the pressure-temperature region of HAT-P-12b's atmosphere at the terminator.  These clouds will have noticeable impact on the transmission spectrum only if these species are enhanced over solar metallically ($>$50x) and have a low sedimentation efficiency resulting in highly vertically extended clouds.  Another possible scenario for producing hazes is through the photochemical destruction of methane.  Photochemistry can drive methane into higher order carbon species such as C$_2$H$_2$, C$_2$H$_4$, and C$_2$H$_6$ which can in principle polymerize into long chained soots or poly-cyclic aromatic hydrocarbons.     Future observations will be needed in order to definitively rule out a cloud free atmosphere and/or to potentially identify the culprit cloud/haze composition.

Although existing Spitzer secondary eclipse observations (Todorov et al. 2013) have only resulted in upper limits, a secure detection of features in the planet's emission spectrum could provide a useful complement to transmission spectroscopy. The recently implemented (Deming 2013) spatial scan mode for WFC3 has the potential to improve the signal-to-noise for HST observations of this planet further testing the flat nature of this spectrum.  Shorter wavelength observations with STIS on HST could also provide confirmation of the presence of clouds.

\section{Acknowledgements}
J.-M.D. acknowledges funding from NASA through the Sagan Exoplanet Fellowship program administered by the NASA Exoplanet Science Institute (NExScI).

\end{document}